\documentclass{article}

\usepackage{arxiv}

\usepackage[utf8]{inputenc} % allow utf-8 input
\usepackage[T1]{fontenc}    % use 8-bit T1 fonts
\usepackage{hyperref}       % hyperlinks
\usepackage{url}            % simple URL typesetting
\usepackage{booktabs}       % professional-quality tables
\usepackage{amsfonts}       % blackboard math symbols
\usepackage{nicefrac}       % compact symbols for 1/2, etc.
\usepackage{microtype}      % microtypography
\usepackage{graphicx}
\usepackage{natbib}
\usepackage{doi}
\usepackage{amsmath}

\title{Enhancing Computational Efficiency in NetLogo: Best Practices for Running Large-Scale Agent-Based Models on AWS and Cloud Infrastructures}

\author{
	Michael A. Duprey \\
	RTI International\\
	\texttt{mduprey@rti.org} \\
	\And
	Georgiy V. Bobashev \\
	RTI International\\
	\texttt{bobashev@rti.org} \\
}
\renewcommand{\headeright}{}
\renewcommand{\undertitle}{}
\renewcommand{\shorttitle}{Enhancing Computational Efficiency in NetLogo}

\hypersetup{
pdftitle={Enhancing Computational Efficiency in NetLogo: Best Practices for Running Large-Scale Agent-Based Models on AWS and Cloud Infrastructures},
pdfsubject={cs.DC, cs.MS},
pdfauthor={Michael A. Duprey, Georgiy V. Bobashev},
pdfkeywords={Agent-Based Modeling, Computational Efficiency, NetLogo, Cloud Computing},
}

\begin{document}
\maketitle

\begin{abstract}
The rising complexity and scale of agent-based models (ABMs) necessitate efficient computational strategies to manage the increasing demand for processing power and memory. This manuscript provides a comprehensive guide to optimizing NetLogo, a widely used platform for ABMs, for running large-scale models on Amazon Web Services (AWS) and other cloud infrastructures. It covers best practices in memory management, Java options, BehaviorSpace execution, and AWS instance selection. By implementing these optimizations and selecting appropriate AWS instances, we achieved a 32\% reduction in computational costs and improved performance consistency. Through a comparative analysis of NetLogo simulations on different AWS instances using the wolf-sheep predation model, we demonstrate the performance gains achievable through these optimizations.
\end{abstract}

% keywords can be removed
\keywords{Agent-Based Modeling \and Computational Efficiency \and NetLogo \and Cloud Computing}

\section{Introduction}

Agent-based models (ABMs) have become increasingly important in various fields, including ecology \citep{grimm2005individual, zhang2020ecology}, economics \citep{axtell2022agent, tesfatsion2006agent}, social sciences \citep{epstein1996growing}, and epidemiology \citep{adams2023examining, cerda2024, JARLAIS2022109573, eubank2004modelling, kerr2021covid}, due to their ability to simulate complex systems of interacting agents. Each agent in an ABM can represent an individual, a group, or a component of a larger system, operating under a set of rules governing its behavior and interactions. This bottom-up approach allows researchers to capture emergent phenomena that arise from the local interactions of agents, which might be difficult or impossible to predict using traditional modeling techniques.

NetLogo \citep{wilensky1999netlogo} is one of the most widely used platforms for building ABMs \citep{railsback2019agent}. Its accessibility, with a user interface, extensions, and extensive library of pre-built models, has made it a popular modeling tool among both novice and experienced modelers. NetLogo is particularly valued for its capacity to handle a wide range of model configurations, from simple educational simulations to research-grade models involving thousands of interacting agents.

However, as the scale and complexity of these models have increased, so too have the computational demands. Large-scale ABMs can require substantial processing power and memory, especially when simulations involve large numbers of agents or complex interaction rules. This is further exacerbated when multiple runs are needed to explore different scenarios or parameter configurations, a common need to ensure robustness of results. Such computational demands can overwhelm standard desktop environments, leading to long run times or system crashes due to memory limitations. For instance, the models employed in \citet{adams2023examining} and \citet{cerda2024} make use of NetLogo; however, due to computational limitations, running these models on a laptop is impractical, necessitating large-scale computational resources.

To address these computational challenges, cloud computing offers a solution for scaling up ABM simulations. Platforms such as Amazon Web Services (AWS), Microsoft Azure, and Google Cloud Platform provide easy access to high-performance computing resources tailored to the needs of the simulation. This flexibility enables researchers to run large-scale simulations more efficiently, both reducing the time required to obtain results and allowing for more complex analyses than would be feasible on a single machine.

While deployment of NetLogo models on cloud infrastructures presents a significant opportunity to enhance the efficiency and scalability of ABMs, maximizing the benefits of this integration requires careful optimization of both the NetLogo environment and the cloud resources. This includes choosing the appropriate version of NetLogo, optimizing Java Virtual Machine parameters, configuring simulation experiments to make use of the cloud platform, and selecting appropriate instance types that align with the specific computational demands of the model. Although NetLogo does not support the Message Passing Interface (MPI), there are a number of other approaches that can facilitate complex models with a large number of simulations.

This manuscript aims to serve as a guide to optimizing NetLogo for large-scale ABMs, with a focus on AWS. Although the emphasis is on AWS, many of the principles discussed are applicable to other cloud platforms.

\section{Optimizing NetLogo}

Optimizing NetLogo for large-scale simulations involves implementing several key strategies to enhance performance. A crucial step is refactoring the model's code to improve efficiency. Comprehensive guidance on this topic is provided by \citet{railsback2017improving}. Several of the authors' recommendations center around methods to reduce or avoid filtering agentsets and strategies such as using state variables instead of links between agents. Furthermore, they suggest employing NetLogo's profiling tools to identify and address performance bottlenecks, allowing modelers to focus optimization efforts where they have the greatest impact.

\subsection{Use of NetLogo 6.4.0+}
We strongly suggest that modelers ensure they are using NetLogo 6.4.0 or newer versions. NetLogo version 6.4.0 \citep{netlogo640release}, released in November 2023, introduced several key improvements aimed at better handling large-scale simulations. These include dynamic memory management adjustments and the resolution of performance issues in BehaviorSpace, NetLogo's tool for running systematic experiments. These updates are particularly important for researchers looking to run extensive simulations, as they help mitigate some common issues such as out-of-memory errors and inefficient data handling during model execution.

The default maximum memory usage is now set to 50\% of the system memory, as opposed to the previous static limit of 1~GB. This change allows NetLogo to utilize available memory more effectively. Users running complex models will likely notice reduced out-of-memory errors, leading to smoother and more reliable simulations. For high-memory systems, allocating an even larger portion to NetLogo can further improve execution speed. This can be done by adjusting the Java options in accordance with the system's specifications.

\subsection{Optimizing Java Options for NetLogo}

NetLogo runs on a Java Virtual Machine (JVM), and optimizing the JVM settings can lead to further improvements in the performance of simulations, particularly when working with large and complex models and on systems with a large amount of available memory. By fine-tuning the Java options, users can manage memory allocation, enhance garbage collection, and optimize thread management, all of which can contribute to efficiency gains. The NetLogo FAQ guide provides guidance on locating the configuration files for Java options used by NetLogo \citep{netlogoFAQModelSize}.

\subsubsection{Memory Allocation}

Proper allocation of memory is one of the most crucial aspects of optimizing Java for NetLogo. By default, the JVM allocates a certain amount of memory to NetLogo, but these default settings might not be sufficient for large-scale models. This is particularly important for simulations involving large agent populations or models that require storing extensive data in memory. Proper memory allocation helps prevent out-of-memory errors, which can cause simulations to crash or perform poorly. The \texttt{-Xmx} option in Java allows setting the maximum heap size -- the memory space where Java objects are created and maintained -- which determines how much memory NetLogo can use during runtime.

For example, if a dedicated AWS instance has 16~GB of RAM, setting the maximum heap size to 12~GB (\texttt{-Xmx12g}) can allow NetLogo to use a large portion of the system's memory while leaving 4~GB of RAM for the operating system and any necessary background processes. A common guideline is to allocate no more than 75–80\% of the total physical memory to the JVM in memory-intensive applications.

In addition to setting the maximum heap size, modelers can also use the \texttt{-Xms} option to set the initial heap size. Consider setting \texttt{-Xms} to a lower value than \texttt{-Xmx} to allow the JVM flexibility in adjusting the heap size. This can help the JVM meet specified pause time goals by expanding the heap if necessary. For example, use \texttt{-Xms8g -Xmx12g} on a system with 16~GB of RAM, allowing the JVM to adjust the heap size between 8~GB and 12~GB.

\subsubsection{Garbage Collection Optimization}

Garbage collection (GC) is the process by which the JVM reclaims memory occupied by objects that are no longer in use. While GC is essential for maintaining memory efficiency, it can introduce pauses in execution, particularly in large simulations. Optimizing garbage collection settings can help mitigate these interruptions.

The default garbage collector in the JVM varies depending on the Java version. For Java 8, the default is the Parallel GC, while in Java 9 and later, the Garbage-First Garbage Collector (G1GC) is the default for server-class machines. While the default garbage collector works well for most models, for large-scale simulations, users might benefit from using the G1GC, which is designed for applications with large heaps. The G1GC (activated with \texttt{-XX:+UseG1GC}, if not by default) divides the heap into regions and performs GC incrementally, which can reduce the length and frequency of GC pauses compared to other collectors like the Parallel GC. This is particularly beneficial for simulations that require consistent runtime performance without significant delays.

Another useful option is \texttt{-XX:MaxGCPauseMillis}, which allows a modeler to specify a soft target for the maximum pause time, in milliseconds, during garbage collection. For instance, setting \texttt{-XX:MaxGCPauseMillis=200} aims to limit GC pauses to 200~ms. If the intervals between garbage collection events are too long, a large amount of unused memory can accumulate on the heap, making subsequent garbage collection processes more costly in terms of time and resources. Conversely, if garbage collection occurs too frequently because the intervals are too short, the application may spend a disproportionate amount of time on memory management tasks rather than executing its primary computational task. We recommend always profiling the model before and after making GC adjustments to ensure that changes lead to actual performance improvements.

For NetLogo simulations, optimizing thread management through JVM options has limited impact due to the application's largely single-threaded execution model. The \texttt{-XX:ParallelGCThreads=<number>} option specifies the number of threads used by the garbage collector, which may improve garbage collection performance in some cases but does not affect the execution of the simulation code itself. Increasing the number of parallel GC threads can reduce garbage collection pause times, particularly if garbage collection is identified as a performance bottleneck through Java profiling tools, such as YourKit.

\subsubsection{Other JVM Options}

Additional JVM options may further optimize NetLogo's performance:
\begin{itemize}
    \item \texttt{-Dfile.encoding=UTF-8}: Sets the default character encoding to UTF-8, ensuring consistent handling of text data across different environments. This option is important for models that process text data or rely on specific character encodings, preventing issues related to character representation.
    \item \texttt{-server}: Forces the JVM to run in server mode, optimizing performance for long-running simulations by enabling advanced compilation and optimization techniques. This option is particularly useful when running models on dedicated servers or cloud instances where long-term performance is prioritized over startup time. Verify whether the JVM defaults to server mode on the system to determine if this option is necessary.
\end{itemize}

\section{Optimizing BehaviorSpace}

BehaviorSpace is NetLogo's built-in tool for running experiments that systematically vary model parameters and collect output data \citep{wilensky1999netlogo}. It allows researchers to perform sensitivity analyses, parameter sweeps, and replicate simulations efficiently. Users define experiments by specifying the parameters to vary, the range of values for each parameter, the number of repetitions per parameter set, and the metrics to record. BehaviorSpace then executes the model iteratively, systematically exploring the defined parameter space and recording the results for analysis. The following strategies can help optimize BehaviorSpace performance, particularly when dealing with complex models or extensive parameter spaces.

\subsection{Parallel Execution of Simulations}

BehaviorSpace supports parallel execution of simulations on multi-core systems and manages the allocation of simulations to different cores automatically. This capability can steer modelers in selecting appropriate cloud computing instances, favoring those with a large number of CPU cores. By default, BehaviorSpace will run using \texttt{floor(0.75 * <number of processors>)}, however, for a dedicated modeling AWS instance, the number of cores may be judiciously increased further. It is important to ensure that the system has sufficient memory and processing resources to handle concurrent simulations without causing resource contention or degradation in performance.

\subsection{Efficient Data Recording}

When configuring BehaviorSpace output, opt for the "table" format rather than "spreadsheet." The table format is more memory-efficient because results are written to disk at the completion of each simulation run, preventing accumulation of data in memory throughout the experiment. Additionally, the table format is easier to process programmatically, allowing for simpler post-simulation analysis. For users who need to aggregate and analyze BehaviorSpace output in table format, an R script is provided as supplementary material to simplify this process. This script reads multiple table output files, aggregates the results, and provides summary statistics, making it easier to interpret large sets of simulation data.

To further optimize data recording, collect only essential metrics (reporters) needed for analysis to minimize the size of output files. Instead of recording data from every agent at every time step, compute and record aggregate statistics such as averages or totals, if appropriate. Adjusting the recording frequency can also reduce data volume; for example, increasing the interval between measurements by computing metrics annually in simulation time or only at the end of the simulation if high-resolution data is not required.

\subsection{Repetitions}

Multiple repetitions per parameter set are essential for capturing the variability inherent in stochastic simulations and for performing statistical analyses. However, increasing the number of repetitions also increases computational load. It is crucial to balance the need for statistical robustness with the available computational resources. Conducting preliminary analyses to determine the minimum number of repetitions required to achieve stable estimates can help optimize this balance.

\subsection{Random Seed Control}

Random seed control is another important factor in ensuring reproducibility and avoiding unnecessary duplication of simulation runs. Using the same random seed across repetitions can lead to identical results, negating the benefits of multiple runs. Conversely, using completely random seeds without control can make it difficult to reproduce results. A practical approach is to set the random seed for each run based on the \texttt{behaviorspace-run-number}, which is a unique identifier that increments with each run in BehaviorSpace. This method ensures that each run uses a different random seed while still allowing for reproducibility; it can be implemented by adding the following to the \texttt{setup} procedure:

\begin{verbatim}
    to setup
      random-seed behaviorspace-run-number
      ; rest of your setup code
    end
\end{verbatim}

\subsection{Running BehaviorSpace Headlessly}
Running BehaviorSpace in headless mode (without the graphical user interface) allows for better resource utilization, particularly in cloud environments. This can be done by using the \texttt{--headless} flag when launching NetLogo from the command line.

The preferred method for running BehaviorSpace headlessly is to call the following from within the NetLogo directory:

\begin{verbatim}
./NetLogo_Console --headless
    --model "<model path>.nlogo"
    --setup-file "<experiment path>.xml"
    --experiment "<experiment name>"
    --threads <thread n>
    --table "<table path>.csv"
\end{verbatim}
 where the setup-file is an exported BehaviorSpace experiment.

Prior to version 6.4.0, an issue existed where BehaviorSpace would update plots even when running headlessly, leading to unnecessary performance overhead \citep{NetLogoIssue2120}. This issue has been addressed in NetLogo 6.4.0 \citep{netlogo640release}. Users running earlier versions should upgrade to avoid this performance issue.

\section{Choosing the Right AWS Instance}

Selecting the appropriate AWS instance type is one of the most important steps in optimizing the performance of large-scale NetLogo simulations. AWS offers a wide range of instance types, each catering to different computational needs, including compute-intensive tasks, memory-heavy workloads, and general-purpose applications. Understanding the characteristics of these instances and matching them to the specific demands of a given NetLogo model can lead to significant improvements in both performance and cost-efficiency.

\subsection{Understanding AWS Instance Types}
AWS instances are categorized based on their hardware configurations, which include varying combinations of CPU, memory, storage, and network capabilities. The focus of our discussion will be on CPU and memory capabilities. The three primary categories relevant to NetLogo simulations are compute-optimized, memory-optimized, and general-purpose instances.

\subsubsection{Compute-Optimized Instances}

Compute-optimized instances, such as the c5 and c6a series in the C family, are ideal for applications that require high processing power relative to memory. They are particularly suited for NetLogo simulations involving intensive calculations or frequent agent updates that do not require extensive memory resources. For example, models that perform complex computations per agent but have a moderate number of agents can benefit from the increased CPU performance of compute-optimized instances.

\subsubsection{Memory-Optimized Instances}

Memory-optimized instances, including the r5 and r6a series in the R family and the x1 series, are designed for applications that require large amounts of memory relative to CPU power. These instances are advantageous for simulations with large agent populations, complex interactions, or models that need to maintain extensive data in memory during runtime. If a NetLogo model experiences frequent memory-related errors or if performance is bottlenecked by memory constraints, a memory-optimized instance is a prudent choice. For instance, simulations that involve large spatial grids or detailed state information for each agent would benefit from this type of instance.

\subsubsection{General-Purpose Instances}

General-purpose instances, such as the m5 and m6a series in the M family and the t3 series in the T family, offer a balanced mix of CPU and memory. These instances are particularly effective when a simulation requires moderate levels of both CPU and memory resources and when cost-efficiency is a priority. They are well-suited to models that do not have extreme resource demands in any single area; for example, exploratory simulations or educational models that do not require maximum performance.

\subsection{Operating System Considerations}

The choice of operating system on an AWS instance could also have some minor impact on the performance of NetLogo simulations. Due to their minimalistic nature, Linux-based instances may reduce non-critical background processes that could consume CPU and memory resources. Linux instances' more obvious benefit is their lower average pricing. Among Linux distributions, Amazon Linux and Ubuntu are likely the best options for NetLogo modeling. Amazon Linux is optimized for performance on AWS and offers tighter integration with AWS services, which could be beneficial if a simulation involves other AWS components, such as S3 for data storage or CloudWatch for monitoring. Ubuntu is widely supported, with a large community and extensive documentation, making it a safe choice for most users.

\subsection{Choosing the Number of Cores for Parallel Processing}

When running NetLogo simulations, especially with BehaviorSpace, the number of CPU cores in an AWS instance significantly impacts overall performance and efficiency through parallel processing.

\subsubsection{Instance Selection Guidelines}

When selecting the number of cores in an AWS instance for parallel processing:

\begin{enumerate}
    \item \textbf{Estimate Total Runs}: Determine the total number of simulation runs and the desired completion time.
    \item \textbf{Calculate Required Cores}: Based on the number of runs and acceptable execution time, estimate the number of cores needed (e.g., to complete 100 runs in 5 hours, with each run taking approximately 5 hours on a single core, you would need at least 20 cores to parallelize the runs effectively).
    \item \textbf{Identify Memory Needs}: Each parallel simulation run consumes memory. Calculate the memory required per run and multiply it by the number of cores used to determine the total memory needed.
    \item \textbf{Consider Instance Types}: Choose an instance type that offers the required number of cores and sufficient memory. Compute-optimized instances like the c5 or c6a series offer high core counts with a balance of memory.
    \item \textbf{Evaluate Cost-Effectiveness}: Higher core counts come with increased costs. Balance the performance benefits against budget constraints.
\end{enumerate}

\subsubsection{Example Scenario}

Suppose your NetLogo model requires 1.5~GB of memory per simulation run, and you wish to run 16 simulations in parallel. You would need an instance with at least 24~GB of memory (1.5~GB $\times$ 16 runs) and 16 CPU cores. An instance like the c5.4xlarge, which offers 16~vCPUs and 32~GB of memory, would be suitable for this scenario.

\subsection{Cost Considerations}
While optimizing for performance is important, it is often equally important to consider the cost implications of running NetLogo simulations on AWS. Instances are typically priced based on the number of vCPUs, memory, storage, and additional services utilized. Compute-optimized instances, such as the c5 and c6a series, generally offer high CPU performance at a lower cost per vCPU, making them ideal for CPU-bound simulations. However, memory-optimized instances, like the r5 and r6a series, provide much higher memory at a premium, which can be necessary for memory-intensive models but may significantly increase overall costs.

\section{Comparative Performance Analysis}
To guide the selection of the most suitable AWS instance type for NetLogo simulations, we conducted a series of comparative performance tests using the wolf-sheep predation model, a standard model in the NetLogo library that simulates predator-prey dynamics \citep{wilensky1997wolf}.

\subsection{Methodology}
To evaluate the performance of different AWS instance types for running NetLogo simulations, we selected three instances: c6a.48xlarge (compute-optimized), m6a.48xlarge (general-purpose), and r6a.48xlarge (memory-optimized). Each instance provided 192 virtual CPUs (vCPUs) but varied significantly in memory capacity and hourly cost (Table~\ref{table1}). All simulations were run using NetLogo version 6.4.0 in headless mode, configured to operate on an Amazon Linux 2023 environment with 64-bit x86 architecture. The storage configuration for each instance was set to an 8 GB gp3 general-purpose SSD. The default Java options were used.

\begin{table}[htbp]
\centering
\caption{AWS Instance Specifications}
\begin{tabular}{|l|c|c|c|}
\hline
\textbf{Instance Type} & \textbf{vCPU} & \textbf{Memory (GB)} & \textbf{Cost per Hour (USD)} \\
\hline
c6a.48xlarge & 192 & 384 & 7.34 \\
m6a.48xlarge & 192 & 768 & 8.29 \\
r6a.48xlarge & 192 & 1,536 & 10.89 \\
\hline
\end{tabular}
\label{table1}
\end{table}

To optimize CPU usage without overloading the system, 163 threads (85\% of the available vCPUs) were allocated for each experiment. This allowed for effective use of parallelism while maintaining system stability and avoiding CPU contention.

The experiments were configured using NetLogo's BehaviorSpace tool to systematically explore the parameter space of the model. BehaviorSpace was configured to systematically vary several model parameters, as detailed in Table~\ref{table2}, to explore the impact of different settings on the simulation results. These parameter sweeps collectively produced a combinatorial total of 41,000 distinct simulation runs. Each experiment was conducted with a single repetition per parameter combination, and the simulations were limited to 1,000 time steps. At the end of each run, key metrics such as the number of sheep, wolves, and patches of grass were recorded.

\begin{table}[htbp]
\centering
\caption{Parameters in BehaviorSpace}
\begin{tabular}{|l|l|}
\hline
\textbf{Parameter}                & \textbf{Range or Value}
\\
\hline
model-version                     & "sheep-wolves-grass" \\
wolf-gain-from-food               & Stepped: 10 to 50 (step size: 1) \\
wolf-reproduce                    & Stepped: 1 to 10 (step size: 1) \\
sheep-gain-from-food              & Stepped: 1 to 10 (step size: 1) \\
sheep-reproduce                   & Stepped: 1 to 10 (step size: 1) \\
initial-number-wolves             & 60 \\
initial-number-sheep              & 1,000 \\
grass-regrowth-time               & 30 \\
\hline
\end{tabular}
\label{table2}
\end{table}

To assess the performance of each AWS instance type and ensure robust statistical results, 10 experiment runs were conducted for each instance type. Key performance metrics, such as runtime, memory usage, CPU utilization, and file system outputs, were logged and analyzed. For each run, system metrics were captured using the standard Unix system resource monitoring utility, \texttt{/usr/bin/time}.

\subsection{Results}

The results of the comparative analysis revealed distinct performance characteristics across the three AWS instance types. The instances were assessed based on their User Time, System Time, Elapsed Time, Max Memory, and Cost Per Experiment, with each metric presented as both a mean and a standard deviation (SD). Below, we present a detailed analysis of the results for each metric.

\begin{figure}[!h]
\caption{Comparison of AWS instance performance and cost-efficiency metrics}
\includegraphics[width=\textwidth]{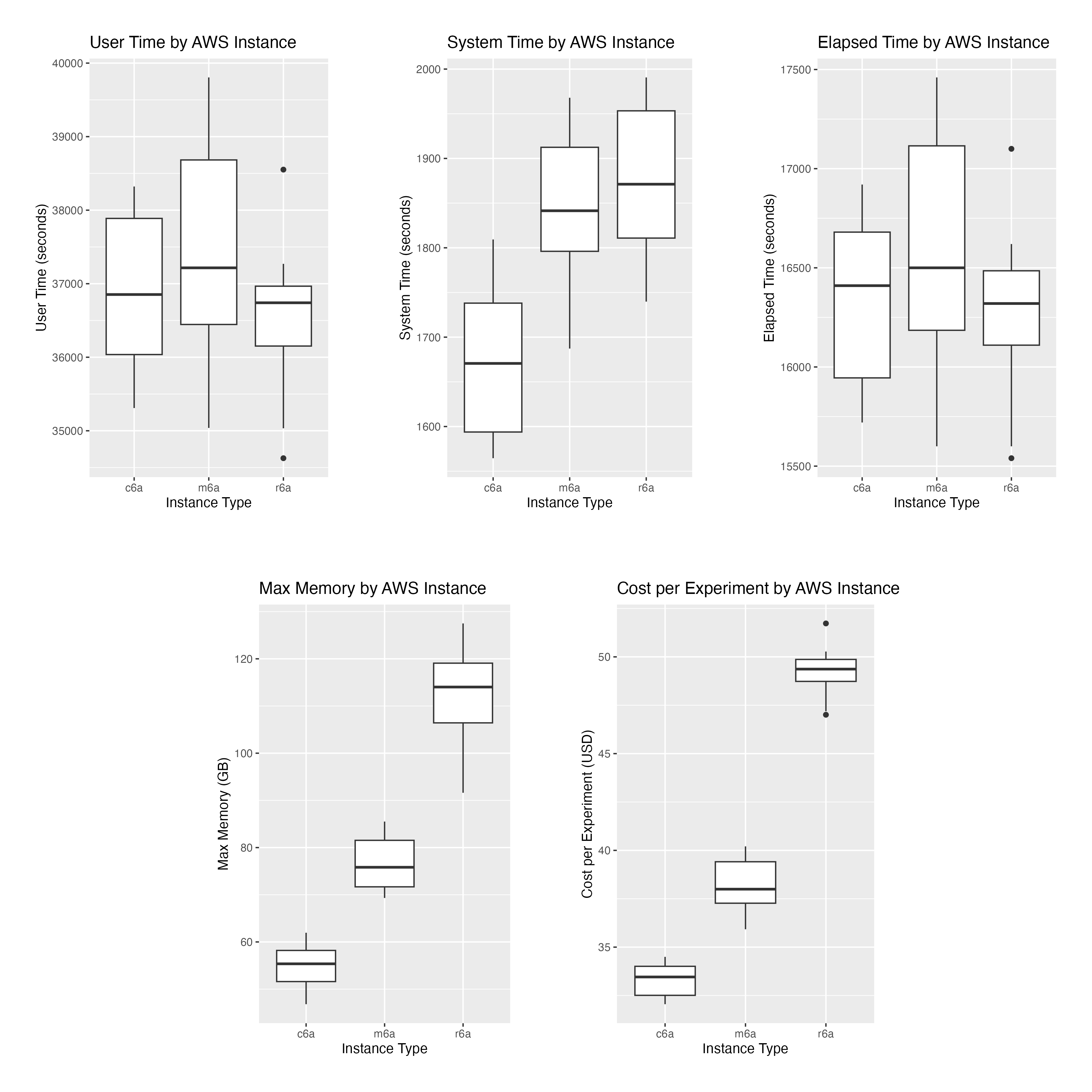}
\centering
\end{figure}

\subsubsection{User Time}

User Time represents the time the CPU spent in user mode, executing tasks. The results show that the m6a instance had the highest Mean User Time at 37,457.75 seconds, followed by c6a at 36,889.18 seconds, and r6a at 36,520.05 seconds. Variability in user time was greatest in the m6a instance (SD = 1649.92 seconds), while c6a exhibited the most consistent performance, with an SD of 1083.97 seconds. These results suggest that while m6a handled slightly more user-level computational tasks, c6a was more reliable in terms of processing time.

\begin{table}[htbp]
\centering
\caption{User Time Statistics for AWS Instances}
\begin{tabular}{|c|c|c|}
\hline
\textbf{Instance} & \textbf{Mean User Time (s)} & \textbf{SD User Time (s)} \\
\hline
c6a & 36,889.18 & 1,083.97 \\
m6a & 37,457.75 & 1,649.92 \\
r6a & 36,520.05 & 1,114.59 \\
\hline
\end{tabular}
\end{table}

\subsubsection{System Time}
System Time refers to the amount of time the CPU spent performing system-level tasks. The r6a instance exhibited the highest Mean System Time at 1876.73 seconds, followed by m6a at 1848.29 seconds, and c6a at 1672.53 seconds. Variability in system time was comparable between m6a (SD = 87.94 seconds) and r6a (SD = 87.08 seconds), while c6a had the lowest variability (SD = 83.93 seconds). The higher system time in r6a is expected, given that memory-optimized instances like r6a often have more intensive system-level resource management.

\begin{table}[htbp]
\centering
\caption{System Time Statistics for AWS Instances}
\begin{tabular}{|c|c|c|}
\hline
\textbf{Instance} & \textbf{Mean System Time (s)} & \textbf{SD System Time (s)} \\
\hline
c6a & 1,672.53 & 83.93 \\
m6a & 1,848.29 & 87.94 \\
r6a & 1,876.73 & 87.08 \\
\hline
\end{tabular}
\end{table}

\subsubsection{Elapsed Time}
Elapsed Time refers to the total time taken to complete the simulations, including all system and user-level tasks. m6a had the longest Mean Elapsed Time at 16,584 seconds, followed by c6a at 16,326 seconds, and r6a at 16,272 seconds. The variability in elapsed time was highest for m6a (SD = 632.58 seconds), indicating that the time required to complete simulations was more variable on this instance. These findings suggest that m6a took more time overall to complete the simulations, while r6a and c6a performed more consistently.

\begin{table}[htbp]
\centering
\caption{Elapsed Time Statistics for AWS Instances}
\begin{tabular}{|c|c|c|}
\hline
\textbf{Instance} & \textbf{Mean Elapsed Time (s)} & \textbf{SD Elapsed Time (s)} \\
\hline
c6a & 16,326 & 434.93 \\
m6a & 16,584 & 632.58 \\
r6a & 16,272 & 462.86 \\
\hline
\end{tabular}
\end{table}

\subsubsection{Max Memory}
Max Memory measures the maximum memory utilized during each run of the simulation. As expected, r6a, the memory-optimized instance, exhibited the highest Mean Max Memory at 112.41 GB, significantly more than m6a at 76.45 GB and c6a at 54.69 GB. The standard deviation for Max Memory was also highest in r6a (SD = 103.64 GB), indicating that memory usage varied significantly across runs on this instance. The high memory usage in r6a aligns with its intended purpose as a memory-optimized instance, although the increased memory consumption did not necessarily correlate with faster task completion times.

\begin{table}[htbp]
\centering
\caption{Max Memory Statistics for AWS Instances (in GB)}
\begin{tabular}{|c|c|c|}
\hline
\textbf{Instance} & \textbf{Mean Max Memory (GB)} & \textbf{SD Max Memory (GB)} \\
\hline
c6a & 54.69 & 4.80 \\
m6a & 76.45 & 5.92 \\
r6a & 112.41 & 103.64 \\
\hline
\end{tabular}
\end{table}

\subsubsection{Cost Per Experiment}
To assess the cost-efficiency of each instance, we calculated the Cost per Experiment based on AWS pricing and the duration of each simulation run. r6a was the most expensive instance, with a Mean Cost per Experiment of \$49.22, followed by m6a at \$38.19 and c6a at \$33.29. The variability in Cost per Experiment was greatest in m6a (SD = \$1.45) and r6a (SD = \$1.40), while c6a had the lowest variability (SD = \$0.89). Given its consistent performance and lower cost, c6a appears to be the most cost-efficient option for these simulations.

\begin{table}[!htbp]
\centering
\caption{Cost per Experiment Statistics for AWS Instances}
\begin{tabular}{|c|c|c|}
\hline
\textbf{Instance} & \textbf{Mean Cost per Experiment (USD)} & \textbf{SD Cost per Experiment (USD)} \\
\hline
c6a & 33.29 & 0.89 \\
m6a & 38.19 & 1.45 \\
r6a & 49.22 & 1.40 \\
\hline
\end{tabular}
\end{table}

\section{Discussion}

Choosing the appropriate AWS instance type is an important step in optimizing both the performance and cost-efficiency of running large-scale NetLogo simulations. Our analysis focused on comparing three instance types---c6a.48xlarge (compute-optimized), r6a.48xlarge (memory-optimized), and m6a.48xlarge (general-purpose)---across several key metrics, including User Time, System Time, Elapsed Time, Max Memory, and Cost per Experiment.

The c6a instance emerged as the most cost-efficient option, delivering consistent performance at the lowest cost per experiment. Specifically, c6a was 32\% more cost-efficient compared to the r6a instance, primarily due to its lower hourly cost while maintaining comparable performance metrics. The relatively low variability in both User Time (SD = 1,084 seconds) and System Time (SD = 84 seconds), combined with its lower cost, suggests that c6a is particularly well-suited for CPU-bound simulations that require stable and predictable processing times. This aligns with the intended purpose of compute-optimized instances, which are designed to maximize CPU performance per dollar spent. For modelers who prioritize computational power and cost control, c6a represents a good trade-off between performance and cost.

In contrast, the r6a instance, while excelling in memory usage, comes at a significantly higher cost. The r6a instance's Max Memory was substantially higher than both c6a and m6a, confirming its suitability for memory-bound workloads. However, despite its memory optimization, r6a did not offer substantial advantages in User Time or Elapsed Time. This discrepancy suggests that for simulations not explicitly constrained by memory, the extra cost associated with r6a is unlikely to be justified. The high memory capacity of r6a may only be fully realized in scenarios where the simulations are large enough to require substantial amounts of RAM, such as when running memory-intensive algorithms, large datasets, or models that generate extensive data.

The m6a instance, designed as a general-purpose option, provides balanced performance across both CPU and memory usage. However, the results indicate that m6a exhibits greater variability in both performance and cost compared to c6a and r6a. This increased variability in User Time and Cost per Experiment may be problematic for modelers who require more predictable performance for simulation workflows. While m6a offers the flexibility to handle a wider range of computational tasks, its higher variability might make it less attractive for simulations where consistency and cost predictability are critical. Thus, m6a may be best suited for use cases where the workload fluctuates between CPU-bound and memory-bound operations but where the user can tolerate some variability in performance.

Our findings also highlight the importance of considering the specific requirements of the simulation when selecting an AWS instance. For modelers focused primarily on cost-efficiency and consistent performance, the c6a instance is an optimal choice. Its lower cost and consistent performance make it ideal for simulations that rely heavily on CPU power but do not require large amounts of memory. On the other hand, the r6a instance may only be warranted for memory-intensive simulations, particularly when the simulation size exceeds the memory capacity of the c6a or m6a instances. However, the higher cost of r6a suggests that its use should be reserved for cases where memory usage is a critical bottleneck and justifies the additional expense.

While our study provides insights into the performance and cost trade-offs of these instance types, several limitations should be acknowledged. First, our experiments were based on a specific and fairly simple NetLogo simulation model (wolf-sheep predation). While this model is a standard benchmark, results may vary depending on the complexity and structure of the simulation. For example, simulations with different computational or memory requirements may yield different performance patterns across these instance types. Additionally, factors such as other concurrent processes, I/O demands, and storage configurations, which were held constant in this study, could further influence the performance of AWS instances in real-world scenarios. Future research could explore how these variables interact with different instance types and impact overall simulation performance.

\section{Conclusions}

This manuscript has outlined best practices for enhancing computational efficiency in NetLogo, particularly when running large-scale agent-based models on AWS. As the complexity and scale of ABMs continue to grow, so too do the demands placed on computational resources. By leveraging the optimization techniques discussed herein, researchers can significantly improve the performance, scalability, and cost-effectiveness of their simulations. Many of the strategies discussed in this manuscript are broadly applicable to cloud platforms other than AWS.

As cloud computing becomes increasingly integral to computational research, understanding how to optimize both software and hardware resources is essential for maximizing the potential of ABMs. The ability to efficiently run large-scale simulations enables researchers to explore more complex systems, conduct more extensive parameter sweeps, and ultimately generate deeper insights into the phenomena being modeled.

Moreover, these optimizations contribute to the broader goals of computational sustainability. By improving the efficiency of simulations, researchers can reduce the time, energy, and cost associated with running large-scale models. This is particularly important in the context of academic and research institutions, where computational resources are often shared and where cost-efficiency is a priority.

\section*{Funding}
Research reported in this publication was supported by the National Institute on Drug Abuse of the National Institutes of Health under award number R01DA047994.

\bibliographystyle{unsrtnat}
\bibliography{references}

@article{adams2023examining,
  title={Examining buprenorphine diversion through a harm reduction lens: an agent-based modeling study},
  author={Adams, Jo{\"e}lla W and Duprey, Michael and Khan, Sazid and Cance, Jessica and Rice, Donald P and Bobashev, Georgiy},
  journal={Harm Reduction Journal},
  volume={20},
  number={1},
  pages={150},
  year={2023},
  publisher={Springer}
}

@article{axtell2022agent,
  title={Agent-based modeling in economics and finance: Past, present, and future},
  author={Axtell, Robert L and Farmer, J Doyne},
  journal={Journal of Economic Literature},
  pages={1--101},
  year={2022},
  publisher={American Economic Association}
}

@article{cerda2024,
  author    = {Magdalenaa Cerdá and Ava D. Hamilton and Ayaz Hyder and Caroline Rutherford and Georgiy Bobashev and Joshua M. Epstein and Erez Hatna and Noa Krawczyk and Nabila El-Bassel and Daniel J. Feaster and Katherine M. Keyes},
  title     = {Simulating the Simultaneous Impact of Medication for Opioid Use Disorder and Naloxone on Opioid Overdose Death in Eight New York Counties},
  journal   = {Epidemiology},
  volume    = {35},
  number    = {3},
  pages     = {418-429},
  year      = {2024},
  month     = {May},
  doi       = {10.1097/EDE.0000000000001703}
}

@book{grimm2005individual,
  title={Individual-based Modeling and Ecology},
  author={Grimm, Volker and Railsback, Steven F.},
  year={2005},
  publisher={Princeton University Press}
}

@book{epstein1996growing,
  title={Growing Artificial Societies: Social Science from the Bottom Up},
  author={Epstein, Joshua M. and Axtell, Robert},
  year={1996},
  publisher={MIT Press}
}

@article{eubank2004modelling,
  title={Modelling disease outbreaks in realistic urban social networks},
  author={Eubank, Stephen and Guclu, Hasan and Kumar, VS Anil and Marathe, Madhav V. and Srinivasan, Aravind and Toroczkai, Zolt{\'a}n and Wang, Nan},
  journal={Nature},
  volume={429},
  number={6988},
  pages={180--184},
  year={2004},
  publisher={Nature Publishing Group}
}

@article{JARLAIS2022109573,
title = {Modeling HIV transmission among persons who inject drugs (PWID) at the “End of the HIV Epidemic” and during the COVID-19 pandemic},
journal = {Drug and Alcohol Dependence},
volume = {238},
pages = {109573},
year = {2022},
issn = {0376-8716},
doi = {https://doi.org/10.1016/j.drugalcdep.2022.109573},
url = {https://www.sciencedirect.com/science/article/pii/S0376871622003106},
author = {Des Jarlais, Don AND Bobashev, Georgiy AND Feelemyer, Jonathan AND McKnight, Courtney}
}

@article{kerr2021covid,
    doi = {10.1371/journal.pcbi.1009149},
    author = {Kerr, Cliff C. AND Stuart, Robyn M. AND Mistry, Dina AND Abeysuriya, Romesh G. AND Rosenfeld, Katherine AND Hart, Gregory R. AND Núñez, Rafael C. AND Cohen, Jamie A. AND Selvaraj, Prashanth AND Hagedorn, Brittany AND George, Lauren AND Jastrzębski, Michał AND Izzo, Amanda S. AND Fowler, Greer AND Palmer, Anna AND Delport, Dominic AND Scott, Nick AND Kelly, Sherrie L. AND Bennette, Caroline S. AND Wagner, Bradley G. AND Chang, Stewart T. AND Oron, Assaf P. AND Wenger, Edward A. AND Panovska-Griffiths, Jasmina AND Famulare, Michael AND Klein, Daniel J.},
    journal = {PLOS Computational Biology},
    publisher = {Public Library of Science},
    title = {Covasim: An agent-based model of COVID-19 dynamics and interventions},
    year = {2021},
    month = {07},
    volume = {17},
    url = {https://doi.org/10.1371/journal.pcbi.1009149},
    pages = {1-32},
    number = {7},

}

@misc{netlogo640release,
  author       = {Wilensky, Uri},
  title        = {NetLogo 6.4.0 Release Notes},
  year         = {2023},
  howpublished = {\url{https://ccl.northwestern.edu/netlogo/docs/versions.html#version-640-november-2023}},
  note         = {Center for Connected Learning and Computer-Based Modeling, Northwestern University. Evanston, IL},
}

@misc{netlogoFAQModelSize,
  author       = {Wilensky, Uri},
  title        = {NetLogo FAQ: How big can my model be? How many turtles, patches, procedures, buttons, and so on can my model contain?},
  year         = {2023},
  howpublished = {\url{http://ccl.northwestern.edu/netlogo/docs/faq.html}},
  note         = {Center for Connected Learning and Computer-Based Modeling, Northwestern University. Evanston, IL},
}

@misc{NetLogoIssue2120,
  author       = {NetLogo Developers},
  title        = {Issue \#2120: BehaviorSpace updates plots when running headlessly},
  howpublished = {\url{https://github.com/NetLogo/NetLogo/issues/2120}},
  year         = {2023},
  note         = {GitHub Issue},
}

@book{railsback2019agent,
  title={Agent-based and individual-based modeling: a practical introduction},
  author={Railsback, Steven F. and Grimm, Volker},
  year={2019},
  publisher={Princeton University Press}
}

@article{railsback2017improving,
  title={Improving Execution Speed of Models Implemented in NetLogo},
  author={Railsback, Steven F. and Lytinen, Steven L. and Jackson, Stephen K.},
  journal={Journal of Artificial Societies and Social Simulation},
  volume={20},
  number={1},
  pages={3},
  year={2017},
  publisher={JASSS}
}

@incollection{tesfatsion2006agent,
  title={Agent-based computational economics: A constructive approach to economic theory},
  author={Tesfatsion, Leigh},
  booktitle={Handbook of Computational Economics},
  volume={2},
  pages={831--880},
  year={2006},
  publisher={Elsevier}
}

@misc{wilensky1997wolf,
  author       = {Wilensky, Uri},
  title        = {{NetLogo Wolf Sheep Predation model}},
  year         = {1997},
  howpublished = {\url{http://ccl.northwestern.edu/netlogo/models/WolfSheepPredation}},
  note         = {Center for Connected Learning and Computer-Based Modeling, Northwestern University, Evanston, IL}
}

@misc{wilensky1999netlogo,
  author       = {Wilensky, Uri},
  title        = {NetLogo},
  year         = {1999},
  howpublished = {\url{http://ccl.northwestern.edu/netlogo/}},
  note         = {Center for Connected Learning and Computer-Based Modeling, Northwestern University. Evanston, IL},
}

@article{zhang2020ecology,
    author = {Zhang, Bo and DeAngelis, Donald L},
    title = "{An overview of agent-based models in plant biology and ecology}",
    journal = {Annals of Botany},
    volume = {126},
    number = {4},
    pages = {539-557},
    year = {2020},
    month = {03},
    issn = {0305-7364},
    doi = {10.1093/aob/mcaa043},
    url = {https://doi.org/10.1093/aob/mcaa043},
    eprint = {https://academic.oup.com/aob/article-pdf/126/4/539/33745089/mcaa043.pdf},
}

\end{document}